\documentclass[12pt,letterpaper]{article}
\pdfoutput=1

\usepackage{jheppub}

\usepackage[T1]{fontenc}
\usepackage{graphicx}

\graphicspath{{Figures/}}

\title{Classical Transitions for Flux Vacua}

\author[a]{J. Tate Deskins}
\author[a,b]{John T. Giblin, Jr}
\author[c]{I-Sheng Yang}

\affiliation[a]{Department of Physics\\ Kenyon College, Gambier, OH 43022, U.S.A.}
\affiliation[b]{Department of Physics\\ Case Western Reserve University, Cleveland, OH 44106, U.S.A.}
\affiliation[c]{ISCAP and Physics Department\\ Columbia University, New York, NY, 10027, U.S.A.}

\emailAdd{deskinsj@kenyon.edu}
\emailAdd{giblinj@kenyon.edu}
\emailAdd{isheng.yang@gmail.com}

\abstract{We present the simplest model for classical transitions in flux vacua. A complex field with a spontaneously broken $U(1)$ symmetry is embedded in $M_2\times S_1$.  We numerically construct different winding number vacua, the vortices interpolating between them, and simulate the collisions of these vortices.  We show that classical transitions are generic at large boosts, independent of whether or not vortices miss each other in the compact $S_1$.}

\begin{document}
\maketitle
\flushbottom

\section{Introduction}

\begin{figure}[htb]
   \centering
   \includegraphics[width=12cm]{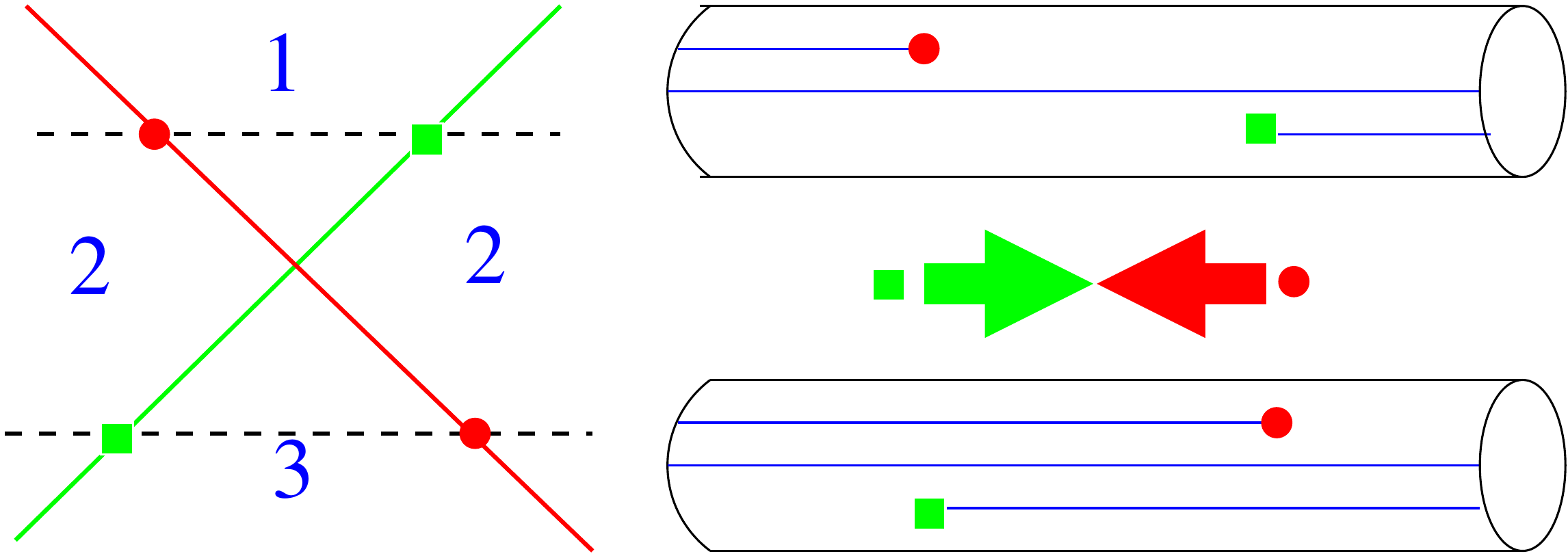} 
   \caption{(left) A conformal diagram showing collision of two domain walls (red and green lines) where the compact extra dimensions is suppressed.  Initially they separate regions containing 3 fluxes and 2 fluxes as labeled.  After the two walls pass through each other, they end up separating regions containing 1 and 2 fluxes.  (right) We chose two time slices, taken at the two times defined by the dashed lines on the conformal diagram, in which we include the compact extra dimension.  The blue flux lines can only originate or terminate at charges.  The red and green domain walls are actually point-like charges, represented by circles and squares respectively.  The charges start in the state defined by the lower tube, representing an early time, then follow the arrow of the corresponding color and evolve to the state defined in the top tube.  It is simply Gauss' law that demands the middle region being left with one flux, resulting in a different compactified vacuum. 
\label{fig-flux}}
\end{figure}

Classical transitions\cite{EasGib09,GibHui10,JohYan10,CliMoo11,KleKri11,YanTye11}---first order phase transitions facilitated by domain wall collisions---have recently been studied in various situations.  One of the original motivations is the simple intuition in \cite{BlaSch09} about flux vacua.  As shown in Fig.~\ref{fig-flux}, different flux vacua have different numbers of field lines threading through the compact extra dimensions.  The boundary between them must contain charges---objects at which field lines can originate or terminate.  If one integrates out the details in the compact dimensions, these boundaries are just domain walls.  However, as localized objects in the extra dimensions, a collision between domain walls does not imply a collision of these charges.  When the charges reside in different places in the extra dimensions, they may simply miss each other.  The natural consequence of this ``missing'' is leaving behind a region with less flux---a transition into a lower vacuum.

There are of course complications to such a simple idea.  For one thing, both charges and flux lines may have strong back reactions on the geometry\cite{MasYan11}.  Even with a fixed background geometry, the charges will have long range interactions with each other.  It is not certain that a clean picture as in Fig.~\ref{fig-flux} is realistic.  

On the other hand, it is known that solitonic objects tend to pass through each other if boosted to large velocities, even if they collide head-on.  This means that classical transitions can take place when vortices collide.  In this paper we present high-resolution simulations that further clarify the dynamics of classical transitions between flux vacua in this model.

\section{Simplest Model of Flux Vacua}

We adopt the simplest model in \cite{BlaSch09}.  Consider a complex scalar field $\phi$ with a spontaneously broken $U(1)$ symmetry,
\begin{equation}
V(\phi) = \frac{\lambda}{4}\left(|\phi|^2-\eta^2\right)^2.
\end{equation}
We embed this theory in a (1+1+1) dimensional spacetime, $M_2\times S_1$,
\begin{equation}
S = \int dt~dx~dy~
\left(\frac{\partial\phi\partial\bar\phi}{2}-V(\phi)\right),
\label{eq-action}
\end{equation}
where $0<y<L$ is the coordinate for the compact extra dimension.  This leads to a discrete family of vacua,
\begin{equation}
\phi_{n,\theta} = \sqrt{\eta^2-\frac{4\pi^2n^2}{\lambda L^2}} 
\exp\left[i\frac{2\pi n y}{L}\right]e^{i\theta}.
\label{eq-vacua}
\end{equation}
Here the integer $n<\frac{\sqrt{\lambda}\eta L}{2\pi}$ is the winding number of the vacuum.  This also defines the energy density spectrum,
\begin{equation}
E_n = 2\frac{\pi^2n^2}{L^2}\left(\eta^2-\frac{\pi^2n^2}{\lambda L^2}\right).
\label{eq-energy}
\end{equation}  
The phase $0<\theta<2\pi$ is an exact symmetry and means nothing for a single vacuum state.  However, when we put two vacua together,
\begin{eqnarray}
\phi(x,y) &=& \phi_{n,\theta_L}, \ \ \ {\rm for} \ \ \ x\rightarrow-\infty,
\nonumber \\
\phi(x,y) &=& \phi_{(n+1),\theta_R}, \ \ \ {\rm for} \ \ \ x\rightarrow\infty,
\end{eqnarray}
the relative phase 
\begin{equation}
\Delta\theta=(\theta_L-\theta_R),
\end{equation}
plays an important role.  The interpolation of minimum energy between these two asymptotic forms must contain a single vortex --- a localized region of high energy density.  Since their phases agree at
\begin{equation}
y=L\frac{\Delta\theta}{2\pi},
\end{equation}
this phase remains constant through out the interpolation.  Therefore, at the opposite side,
\begin{equation}
y=L\frac{\Delta\theta\pm\pi}{2\pi},
\end{equation}
the phase must go through the most dramatic change.  That is the center of the vortex.

This vortex will be attracted to the $(n+1)$th vacuum, and eventually relax the entire region into the lower vacuum.  If we integrate out the compact dimension $y$, we describe this is a domain wall in the $x-t$ spacetime that moves to engulf the region of higher vacuum energy.  To produce collisions of domain walls, we can setup the initial condition with three regions,
\begin{equation}
\begin{array}{ccl}
\phi_{n,-\frac{\Delta\theta}{2}} &\,\,&  {\rm left},  \\
\phi_{(n+1),0} & \,\,& {\rm middle}, \label{eq-setup3} \\
\phi_{n,\frac{\Delta\theta}{2}} &\,\,&  {\rm right}. 
\end{array}
\end{equation}
From the $x-t$ point of view, it is a shrinking false vacuum region with two domain walls that slam into each other.  However when $\Delta\theta\neq0$, it seems like the two vortices will not collide.  This is exactly what the pictorial understanding we gain from Fig.~\ref{fig-flux} leads us to believe; hence, we might expect classical transitions to occur whenever the localized vortices do not collide.

However, we should be careful before exercising such intuition.  Saying that the vortices miss each other implies that they are much smaller than $L$.  It is well known that in a non-compact 2D space, the vortex of this global $U(1)$ does not have a finite size.  That fact of course remains after we compactify one of the spatial dimensions.  Although the center of two vortices may have a large $y$ separation, their profiles still overlap significantly during the collision.

On the other hand, overlapping profiles does not guarantee failure of classical transitions.  In\cite{EasGib09,GibHui10} there was no extra dimensions and the domain walls simply collide with each other.  It was shown that when the collision is very energetic, the field profiles are so highly boosted that they cannot efficiently interact, therefore must remain intact. 

Given these two motivating reasons to believe classical transitions should occur in our model, one might expect that large separation, $\Delta \theta$, or highly boosted domain walls will lead to classical transitions.   At the same time, there should be a low boost and/or small separation regieme where it fails.  The result turns out to be more subtle, as we shall demonstrate later.

\section{Relaxing to the Wall}
\label{sec-relax}
The first challenge that we need to address is how to set up the initial conditions.  The functional form for the field profile that interpolates between local minima of a potential \cite{Col77,CalCol77,ColDel80} is well known for some potentials.  In some cases these initial conditions are analytic; in others relaxation to an approximate solution is necessary before simulations can be run.  In the model presented here, our vacuum states are defined by field configurations in the compact dimension and the interpolation between those states must be found numerically\cite{AguJoh09a,AhlGre10} by relaxing the field to the instanton solution.

The vacuum states of our model are defined by Eq.~(\ref{eq-vacua}).  Our next goal is to numerically setup the situation in Eq.~(\ref{eq-setup3}) that will lead to domain wall collisions.  For numerical reasons, even the uncompactified dimension $x$ can only have a finite size.  We will take that to be $-L_x/2 < x< L_x/2$ and enforce fixed boundary conditions at $x=\pm L_x/2$.  One should not confuse the finite box size $L_x$ with the size of compact extra dimensions $L$ in Eq.~(\ref{eq-vacua}).  We will maintain that $L_x\gg L$ through out the paper. 

\begin{figure}[htbp] 
   \centering
\includegraphics[width=2in]{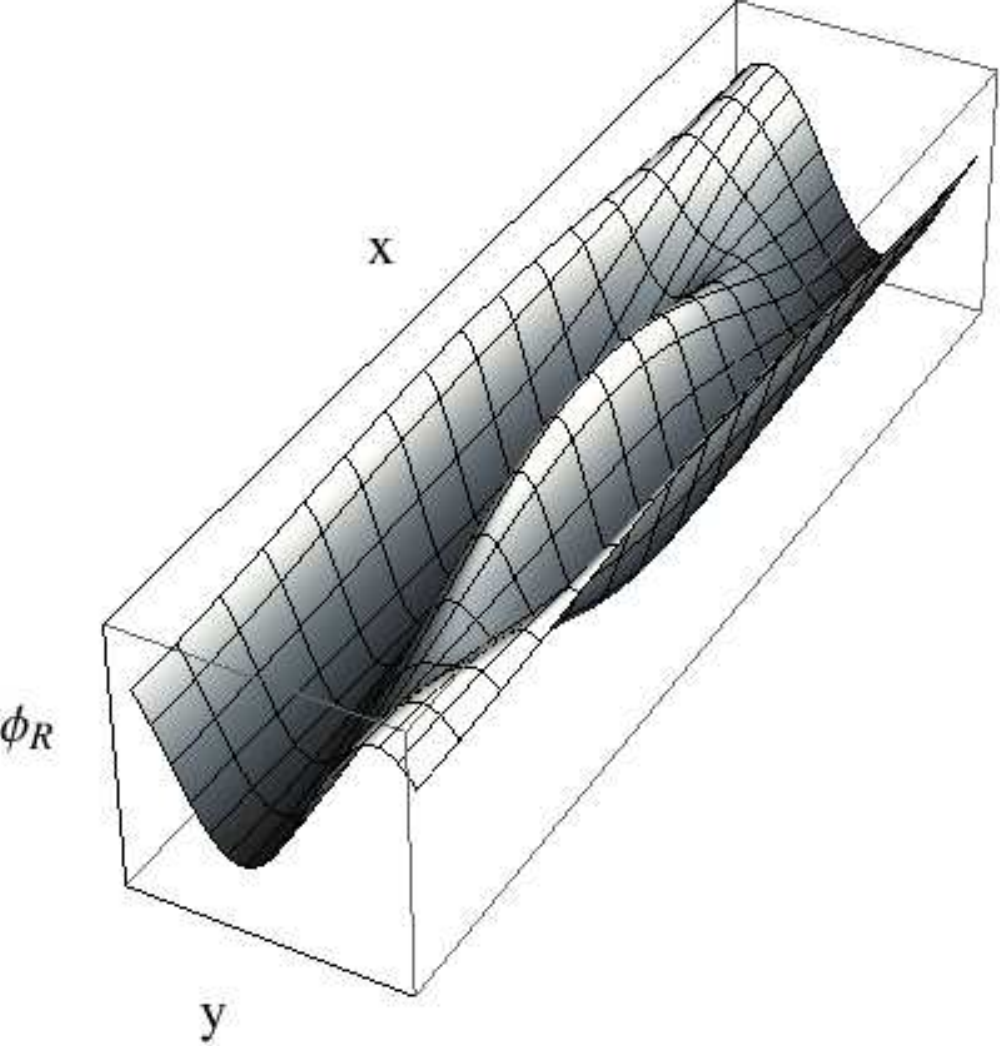}  \includegraphics[width=2in]{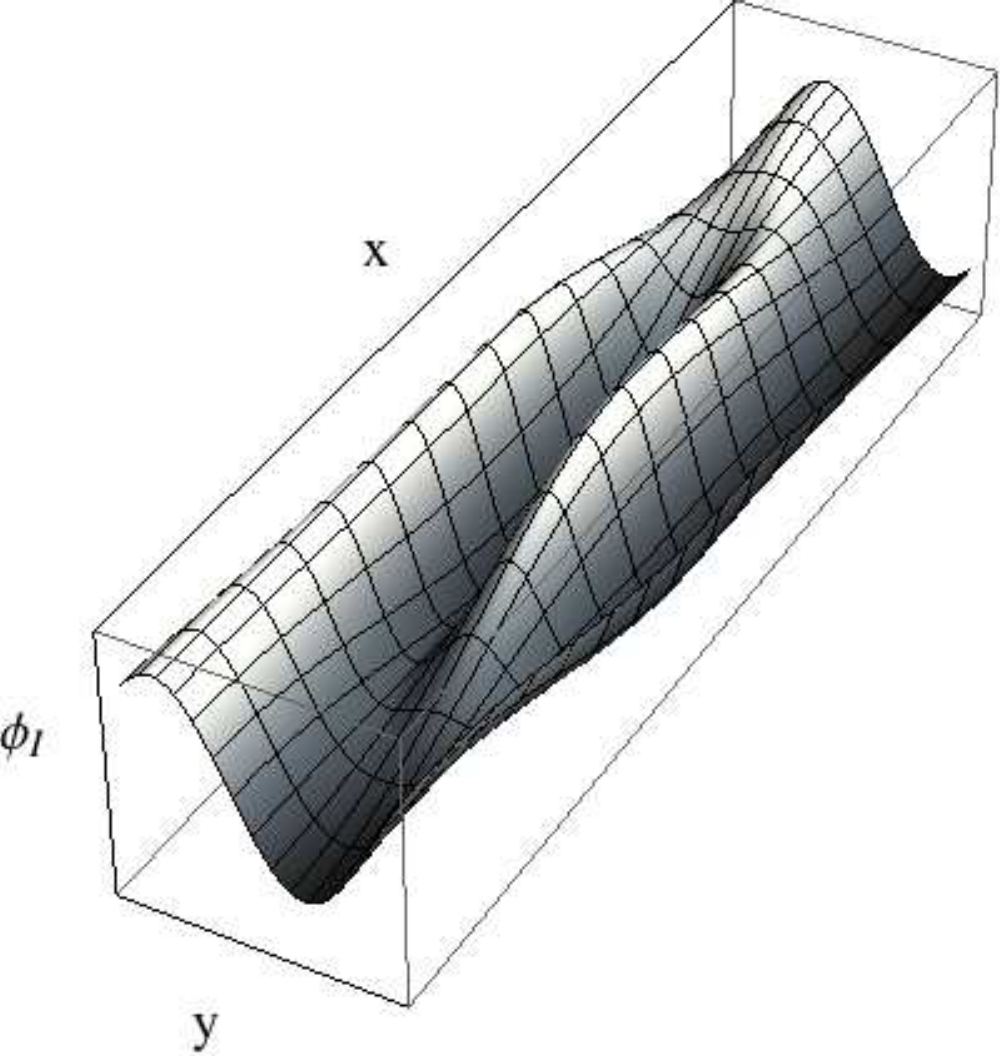}  \includegraphics[width=2in]{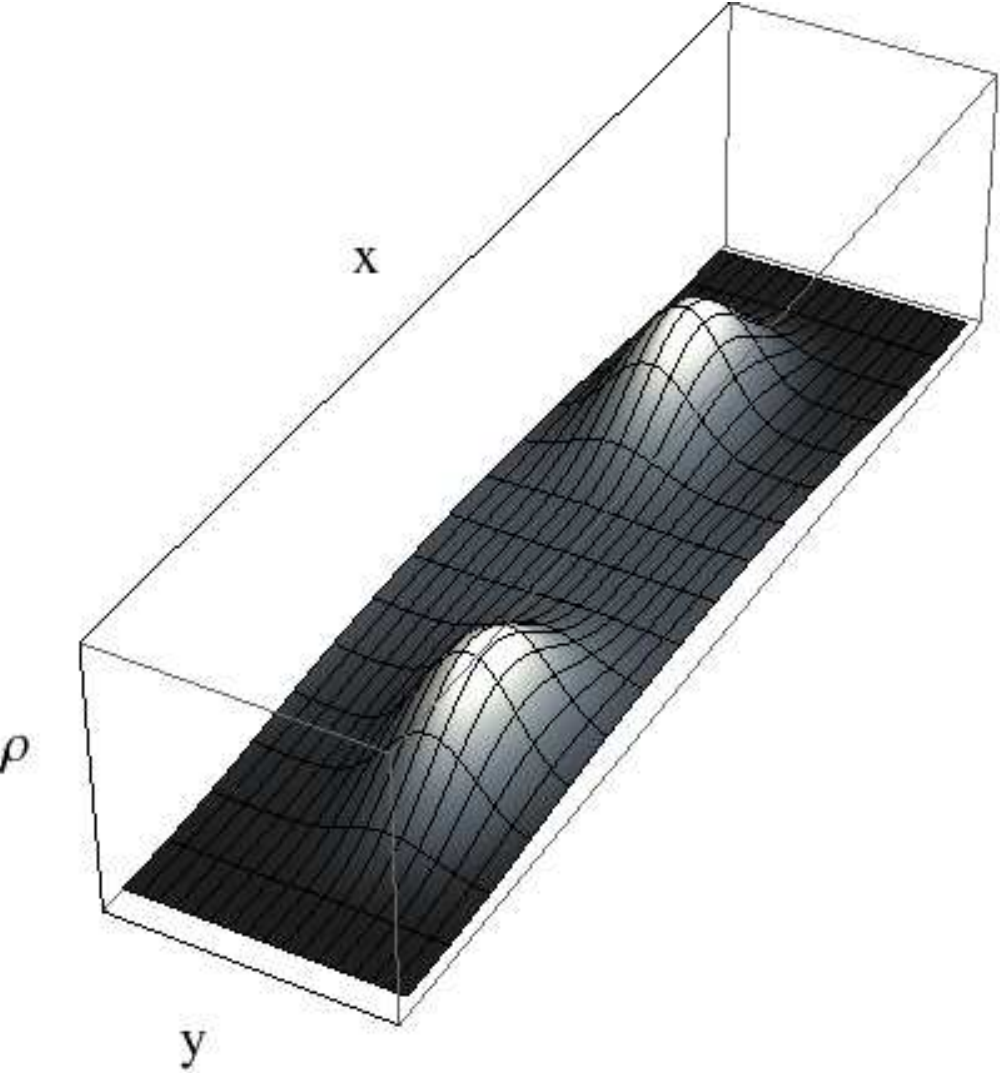} 
   \caption{(left) The real, $\phi_R$, and (middle) imaginary, $\phi_R$, parts of the initial field profile.  (right) The energy density of the initial field profile.}
   \label{initial}
\end{figure}

The initial field profile is a smooth interpolation version of Eq.~(\ref{eq-setup3}) with $n=1$.
\begin{equation}
\phi (t=0) = \left\{
\begin{array}{cc}
\cos^2\left(\frac{\pi}{L_x} x\right)\phi_{2,0} + \sin^2\left(\frac{\pi}{L_x} x \right)\phi_{1,\frac{\Delta\theta}{2}} & -L_x/2<x<0 \\
\cos^2\left(\frac{\pi}{L_x} x\right)\phi_{2,0}+ \sin^2\left(\frac{\pi}{L_x} x\right)\phi_{1,-\frac{\Delta\theta}{2}} & 0<x<L_x/2 
\end{array}
\right. .
\label{eq-setup}
\end{equation}
Fig.~\ref{initial} shows the real and imaginary parts of Eq.~(\ref{eq-setup}) along with a plot of the energy density.

\begin{figure}[htbp] 
   \centering
\includegraphics[width=2in]{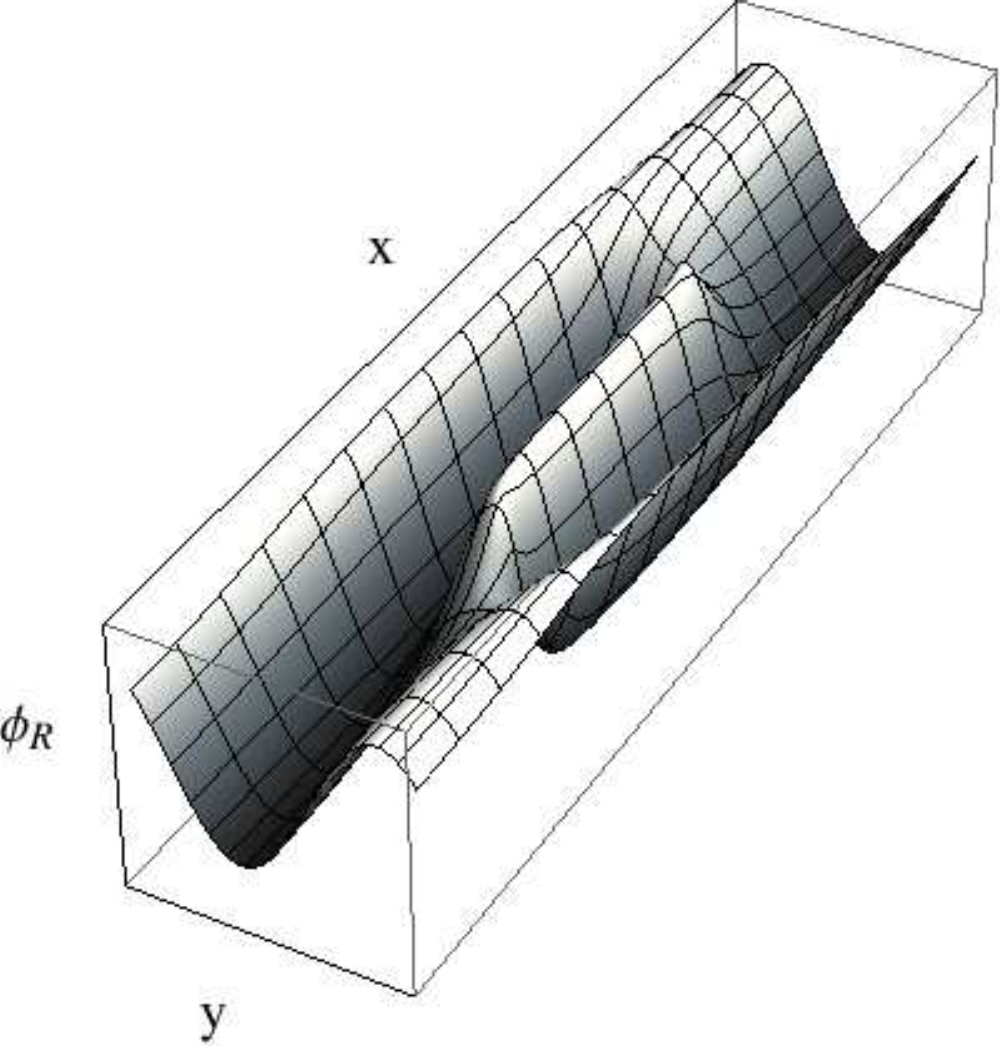}  \includegraphics[width=2in]{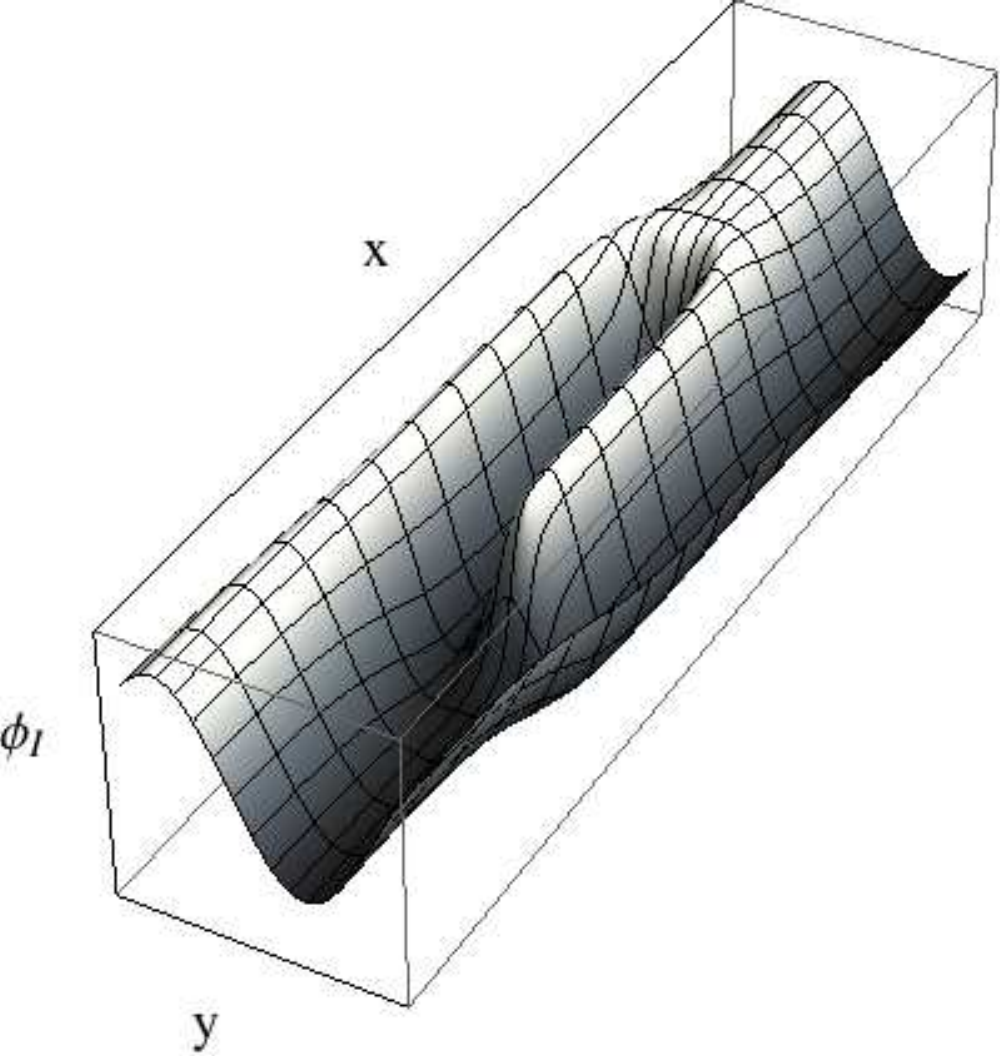}  \includegraphics[width=2in]{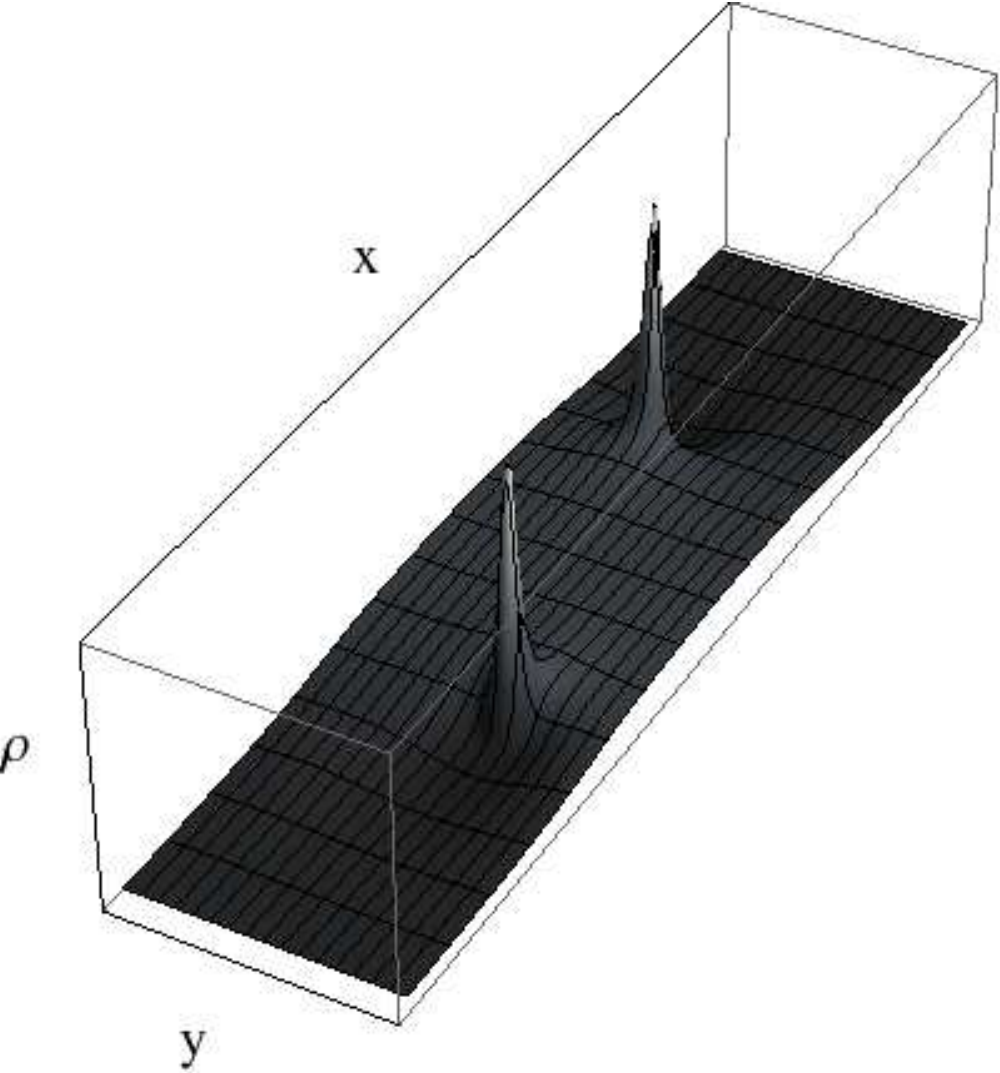} 
   \caption{(left) The real, $\phi_R$, and (middle) imaginary, $\phi_R$, parts of the field profile after relaxation, and also the energy density.  We can see from the energy density plot (right) that it is highly concentrated at two points, which are basically the center of vortices.  We can also see the three vacuum regions in which the energy density is almost constant, and the small difference in energy between the middle region and the ends.}
   \label{start}
\end{figure}

Relaxation is achieved by imposing a period of high friction at the start of the simulation which damps out spurious radiation produced by the field as it relaxes to the instanton solution.  The field values for the majority of the volume of the simulation should be very close to the vacuum states, Eq.~(\ref{eq-vacua}), with localized, although not infinitely so, vortices along the domain walls that separate the different states.  The $y$ separation of the two vortices should follows from our setup of $\Delta\theta$.  This is indeed what we see in Fig.~\ref{start}.  These define the initial conditions for the next step: acceleration and collision.  

Note that, in principle, the motion we hope to see is already happening during this period of high friction.  We are just exploiting a separation of scales.  The eventual accelerations of vortices happens in a longer time scale, comparing to the energy damping from Fig.~\ref{initial} to Fig.~\ref{start}.  In practice, we execute both relaxation to the initial condition and the evolution afterward in one numerical process.  The simulation begins at $t=0$, and reaches Fig.~\ref{start} at $t=t_s$, at which point the friction is turned off (or greatly reduced).

\begin{figure}[htbp] 
   \centering
   \includegraphics[width=3in]{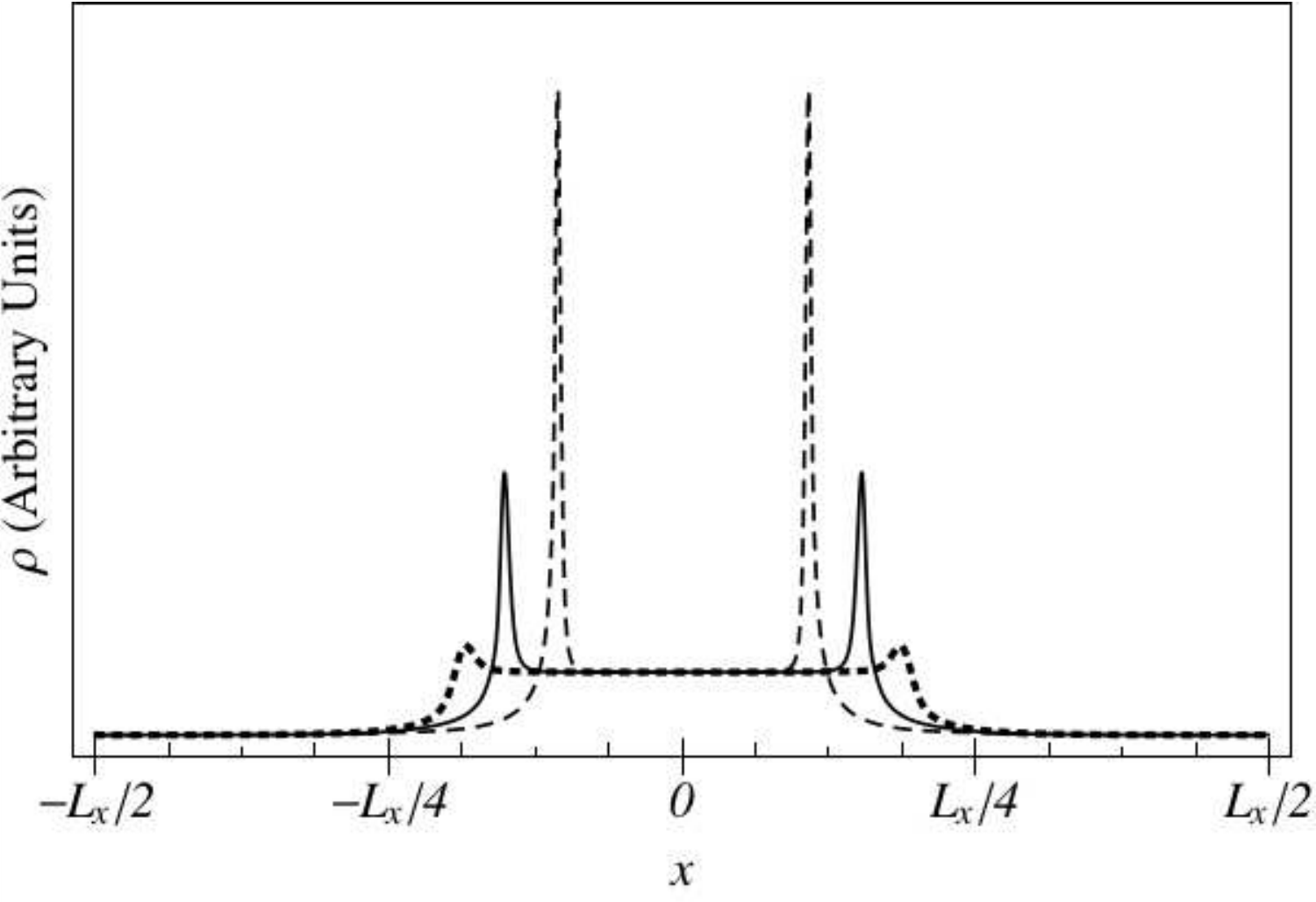}  
   \caption{The integrated energy density as a function of $x$.  The thick-short dashed curve is at $t=t_s$.  The solid and dashed curves are at subsequent times.}
   \label{walls}
\end{figure}

We shall also introduce the ``Einstein Frame'' point of view, in which the details in the compact dimension $y$ are integrated out.  Fig.~\ref{walls} shows the energy density integrated over the compact dimension.  At time $t_s$ we see that the high peaks in Fig.~\ref{start} are actually pretty low energy barriers between the two vacua.  That is because the $n=2$ state in the middle is the highest vacuum in this parameter choice.  We also present the curves at later times when the walls pick up some velocity.  We can see the integrated energy density of the wall acquires a Lorentz boost and grows in magnitude.  This confirms that there is indeed some ``object'' being accelerated during the simulation.

\section{Collisions}

After turning off friction at $t_s$, the field continues to evolve according to the equation of motion one can derive from the action in Eq.~(\ref{eq-action}).  In practice we still keep a small non-zero friction to control numerical noises. As long as the resulting energy loss is tiny compared to the energy scale involve in the simulation, we claim that our result is a good approximation to the exact equation of motion.

Before the collision, as shown in Fig.~\ref{walls}, the vacuum regions and domain walls are quite sharply recognizable just from the energy plot.  However during the collision the non-linear dynamics of the field make energy plots less informative.  We need another quantity to keep track of the dynamics.  Recall that different vacuum states, Eq.~(\ref{eq-vacua}) has different winding numbers.  So we can use the general definition of winding number,
\begin{equation}
w= \frac{1}{2\pi} \oint_C \frac{{\rm Im} (\phi^* d\phi)}{\phi^*\phi} = \frac{1}{2\pi}\oint_{C}\frac{\phi_R}{\left|\phi\right|^{2}}d\phi_I-\frac{\phi_I}{\left|\phi\right|^{2}}d\phi_R~,
\label{eq-winding}
\end{equation}
to recognize different states.  Take the closed contour $C$ in Eq.~(\ref{eq-winding}) be a path along the compact dimension $y$, every $x$ position has a well-defined value of $w$ unless a vortex is exactly there.  Keeping track of the value of $w$ at $x=0$, and evaluating it at the end of the simluation, allows us to tell whether a classical transition has happened.

\begin{figure}[htbp] 
   \centering
   \includegraphics[width=3in]{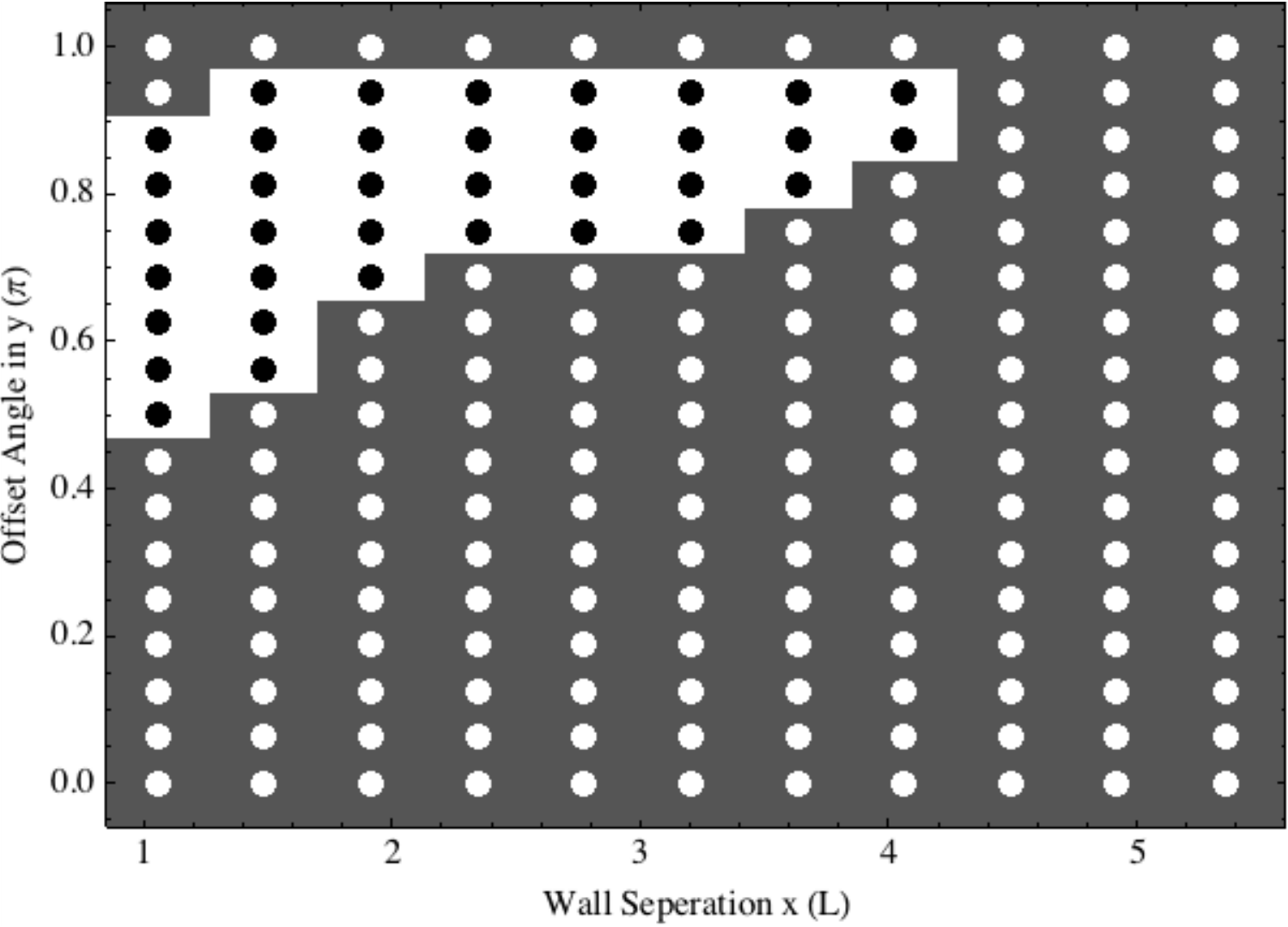}  
   \caption{Contour plot showing the final state of the simulation at $x=0$.  White dots (on grey) represent simulations in which the stable transition to $n=0$ occurred at $x=0$.  Black dots (on white) represent regions that did not transition to the $n=0$ state, and the final state at $x=0$ was $n=1$.}
   \label{fig-scan}
\end{figure}

We are interested in whether the existence of a classical transition depends on the collision energy, and the distances between vortices in the $y$ dimension.  The later can be controlled by $\Delta\theta$ in the initial conditions, Eq.~(\ref{eq-setup}), as explained in Sec.~\ref{sec-relax}.  The collision energy is controlled by the wall separation at $t_s$, since they basically come from the potential energy in the middle region.  We adjust this by changing $L_x$ without changing the definition of the initial profile in Eq.~(\ref{eq-setup}).  The resulting wall separation can be evaluated numerically, and it is accurately proportional to $L_x$.  Scanning through $\Delta\theta$ and wall separation, we present the results in Fig.~\ref{fig-scan}.

There are several features to point out in Fig.~\ref{fig-scan}.  First, as the collision energy increases, eventually classical transition happens for all $\Delta\theta$.  Actually for any fixed $\Delta\theta$, increasing collision energy always helps\footnote{With one exception when the wall separation is close to $L$.  It is unclear what happens there.  When the wall separation is comparable to the compact dimension size, our usual intuition might be misleading.  For example it might not be true that larger separation means more collision energy there.}.  Most interestingly, for a fixed wall separation, increasing $\Delta\theta$ does not really help.  Actually it is the opposite.  All the head-on collisions lead to classical transitions.  It starts to fail at some large $\Delta\theta$, and eventually recovers at the maximum $\Delta\theta=\pi$.

\begin{figure}[htbp] 
   \centering
   \includegraphics[width=2in]{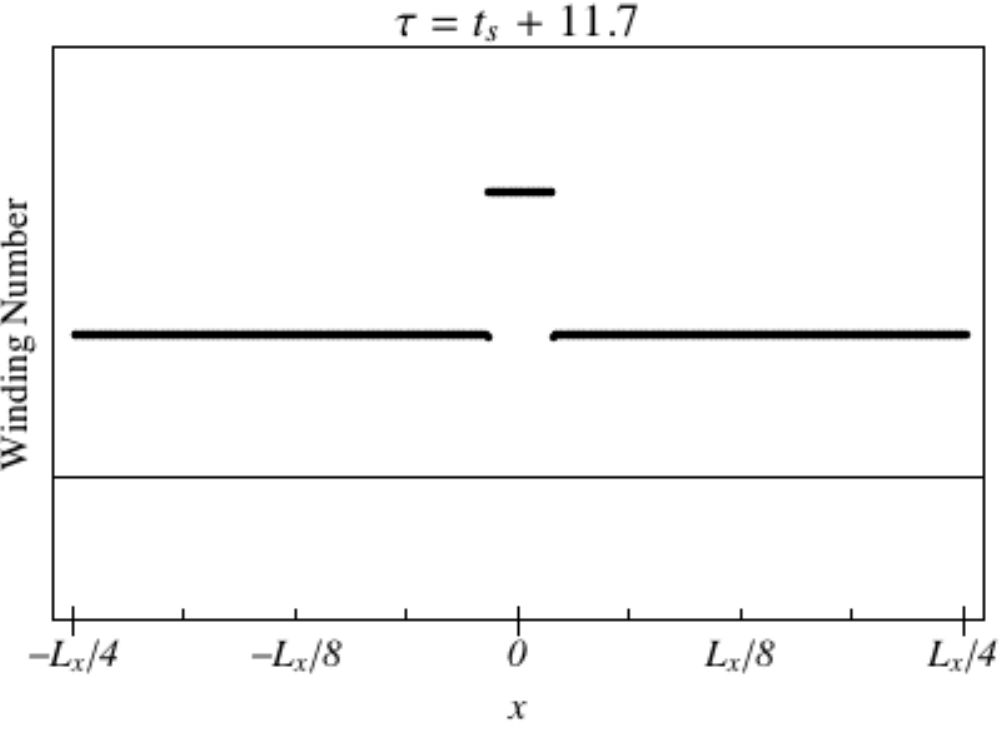}  \includegraphics[width=2in]{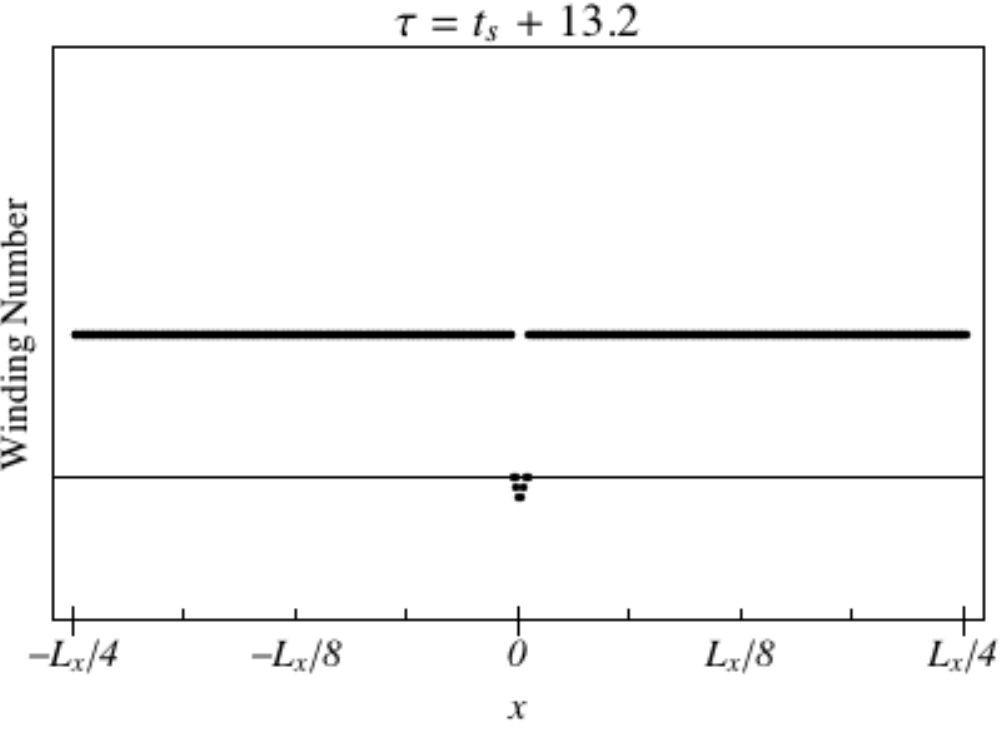}  \includegraphics[width=2in]{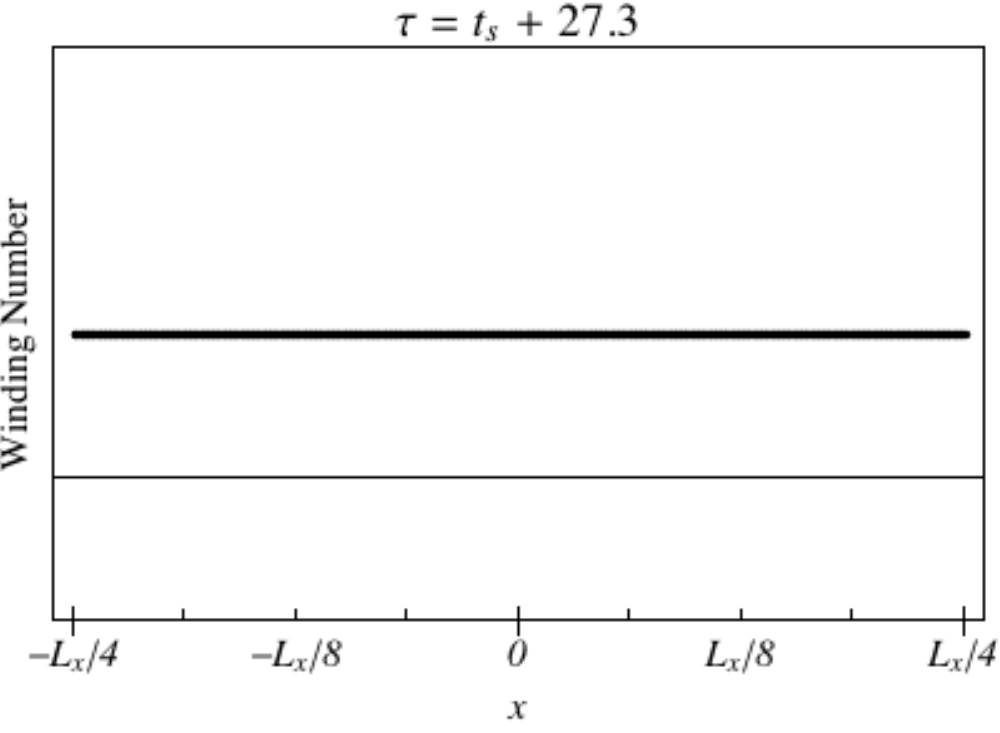} 
\caption{$w(x)$ just before the collision (left), during the collision (middle) and the final state after the collision (right).  This is for wall separation $\approx 1.5L$ and $\Delta\theta=9\pi/16$.}
   \label{fig-4wind}
\end{figure}

\begin{figure}[htbp] 
   \centering
   \includegraphics[width=2in]{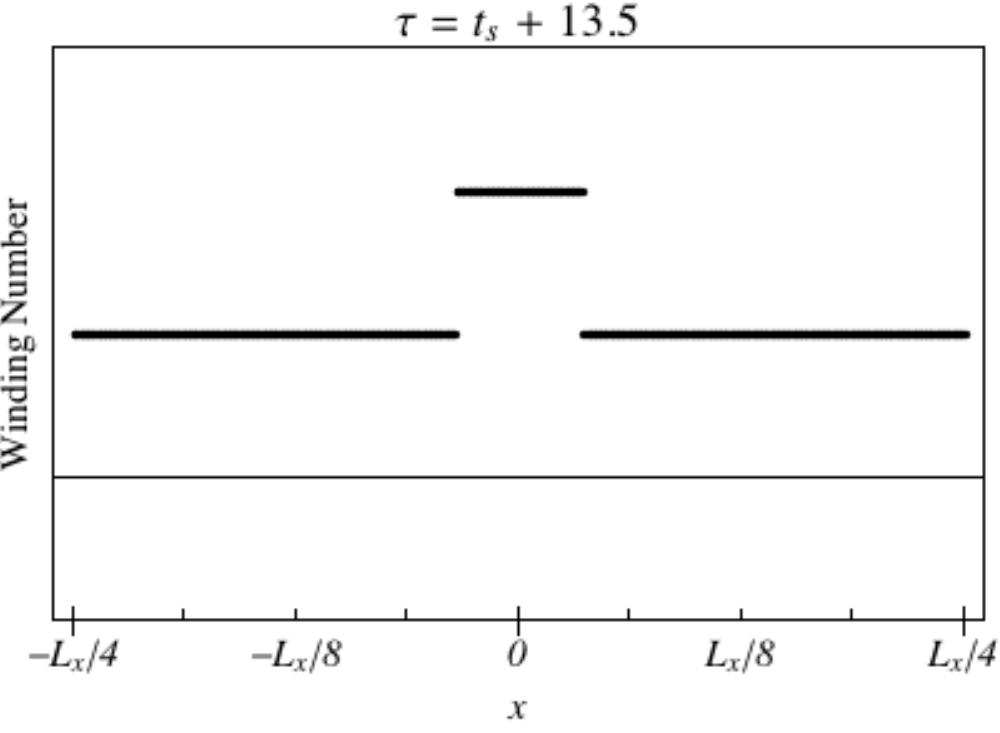}  
 \includegraphics[width=2in]{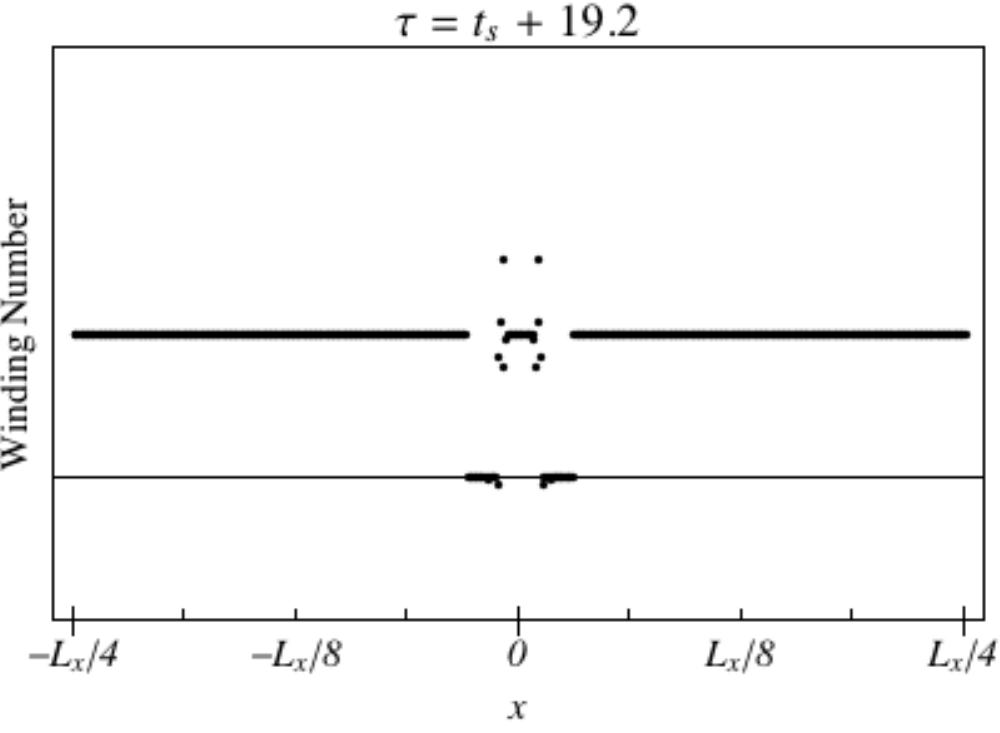}  \includegraphics[width=2in]{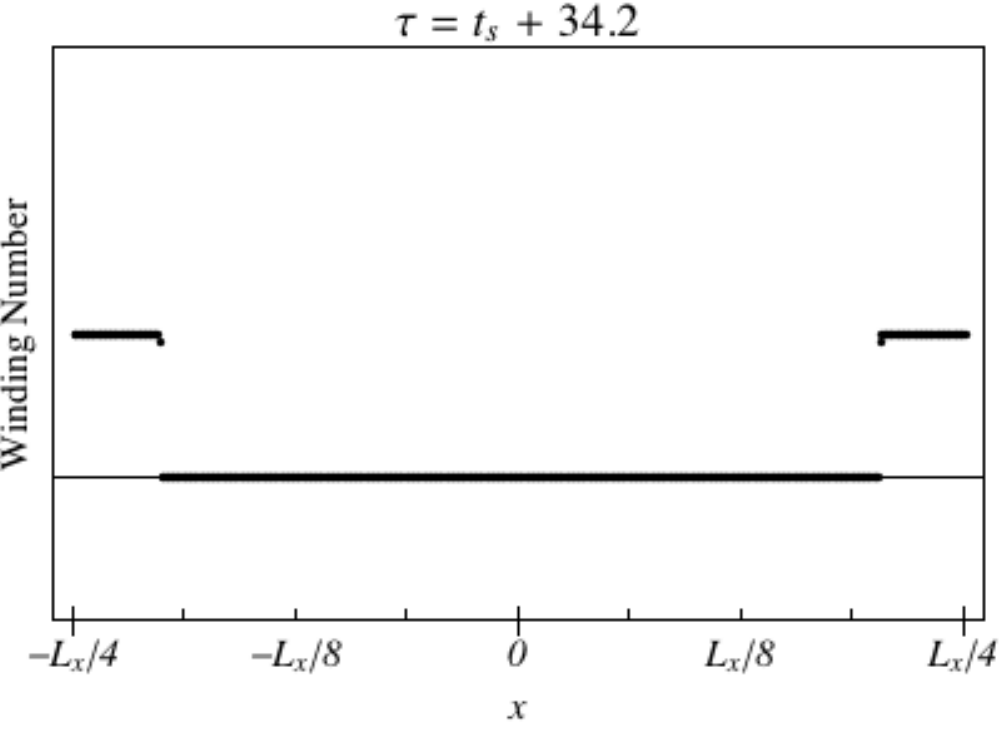} 
 \caption{The similar set of three snapshots of $w(x)$, for wall separation $\approx1.9 L$ and $\Delta\theta=9\pi/16$.}
   \label{fig-5wind}
\end{figure}

We select a few examples to show the winding number plots during the simulations.
To understand the effect of changing the collision energy, we keep $\Delta\theta=9\pi/16$ and show the case of wall separation $x\approx 1.5L$ in Fig.~\ref{fig-4wind}, and $\approx1.9L$ in Fig.~\ref{fig-5wind}.  That is the difference between making the transition or not.  In Fig.~\ref{fig-4wind}, $w(0)$ does dip down to $0$ for a moment, but we did not get two stable vortices flying apart to maintain and expand that region.  At a certain moment in Fig.~\ref{fig-5wind} we actually have 4 vortices separating the entire simulation range into 5 regions.  After that the middle $n=1$ region collapses and gives us a successful transition.

\begin{figure}[htbp] 
   \centering
 \includegraphics[width=2in]{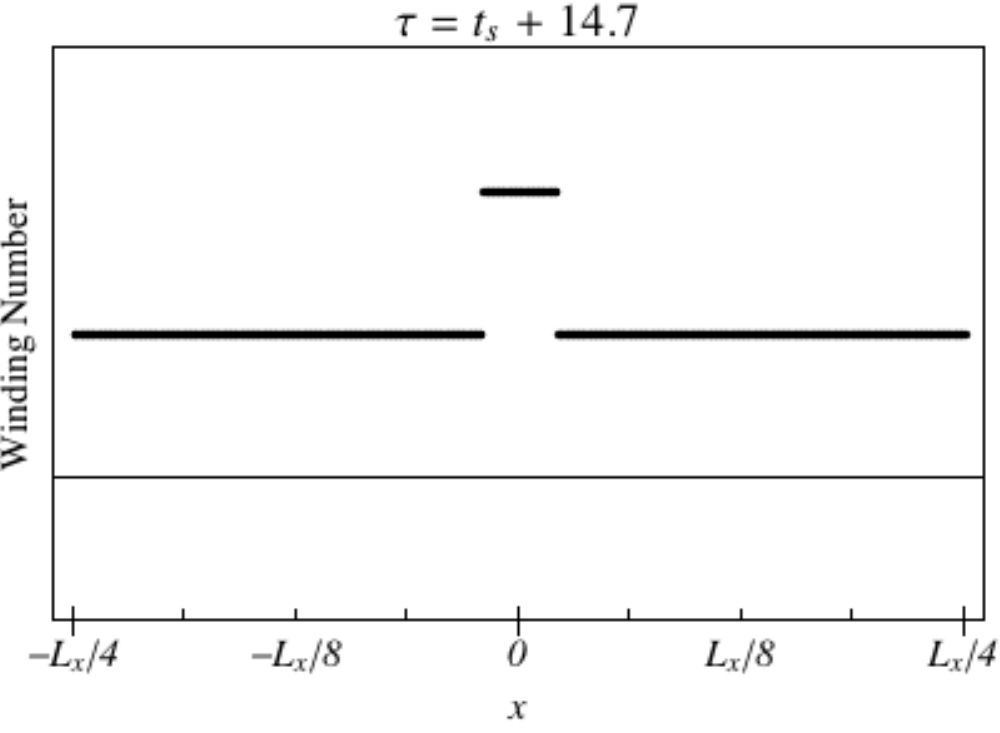}  
 \includegraphics[width=2in]{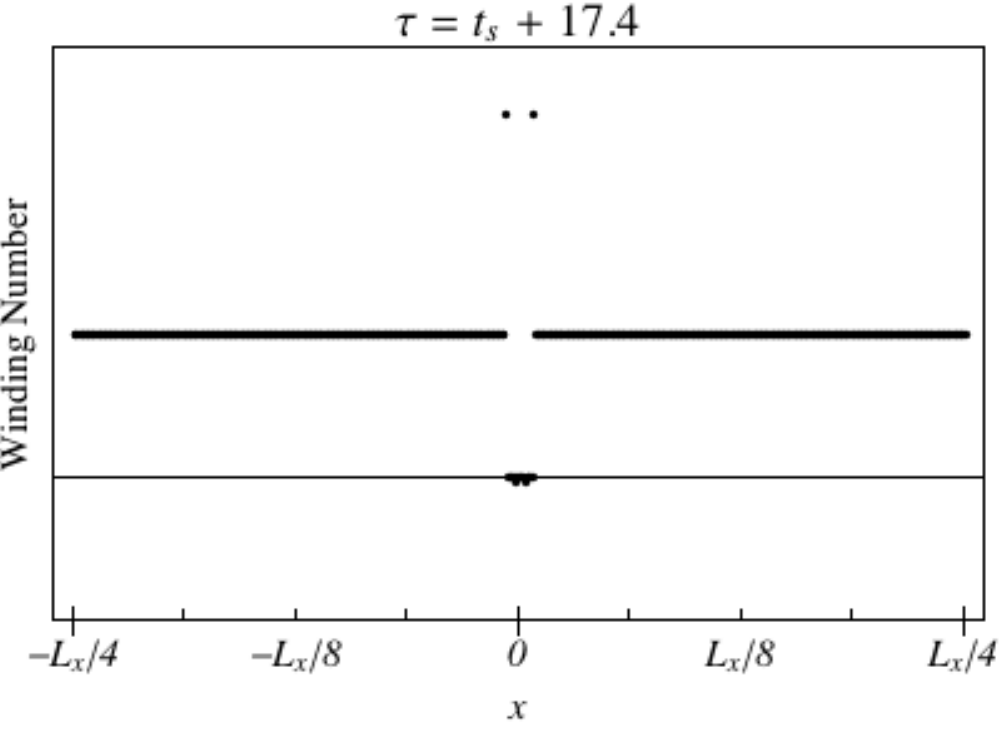}  \includegraphics[width=2in]{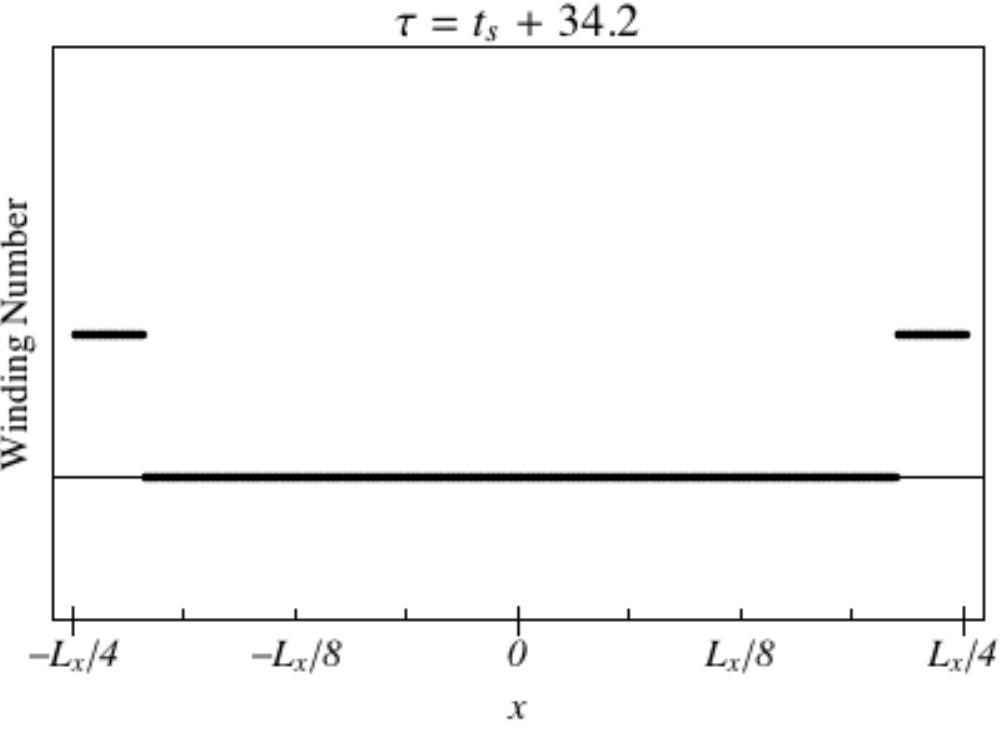} \\
  \includegraphics[width=2in]{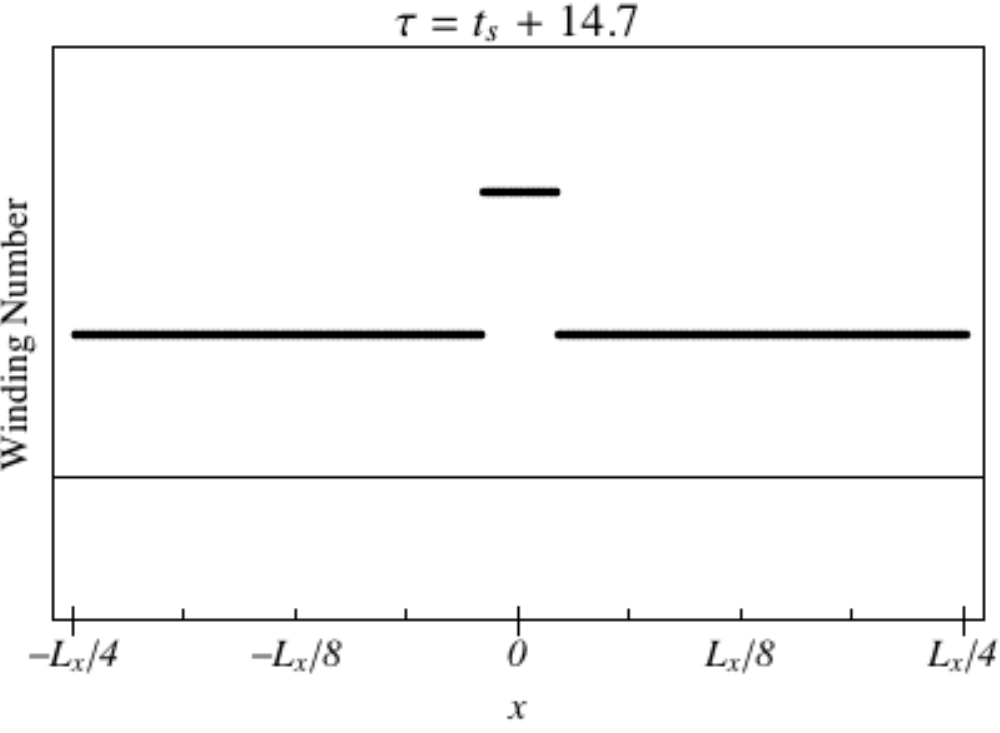}  
 \includegraphics[width=2in]{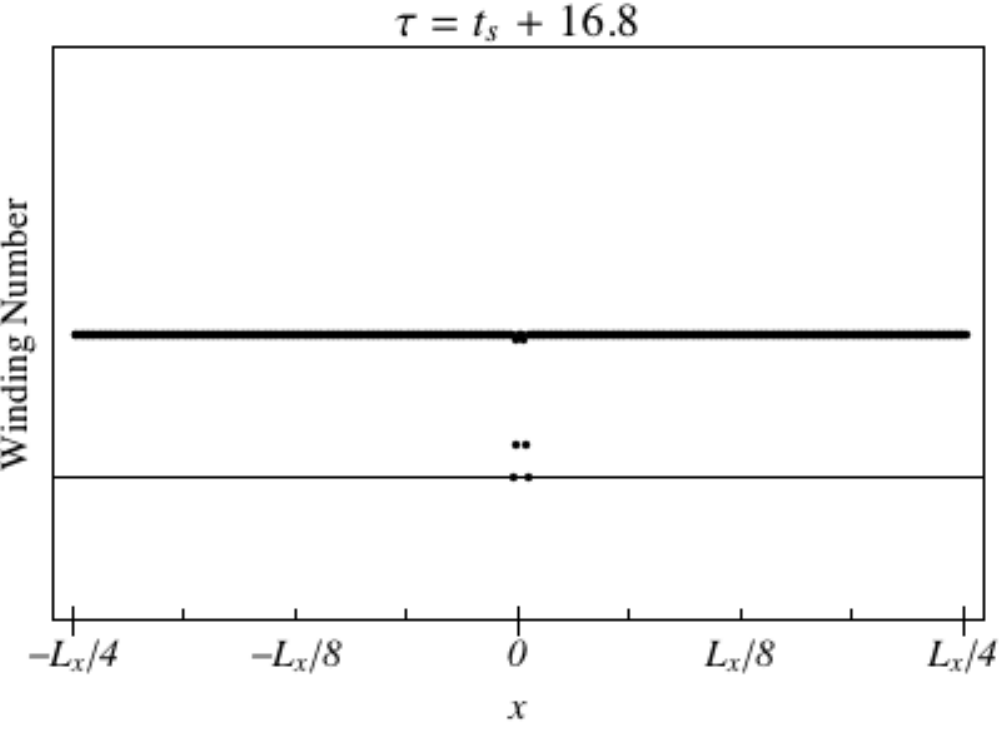}  \includegraphics[width=2in]{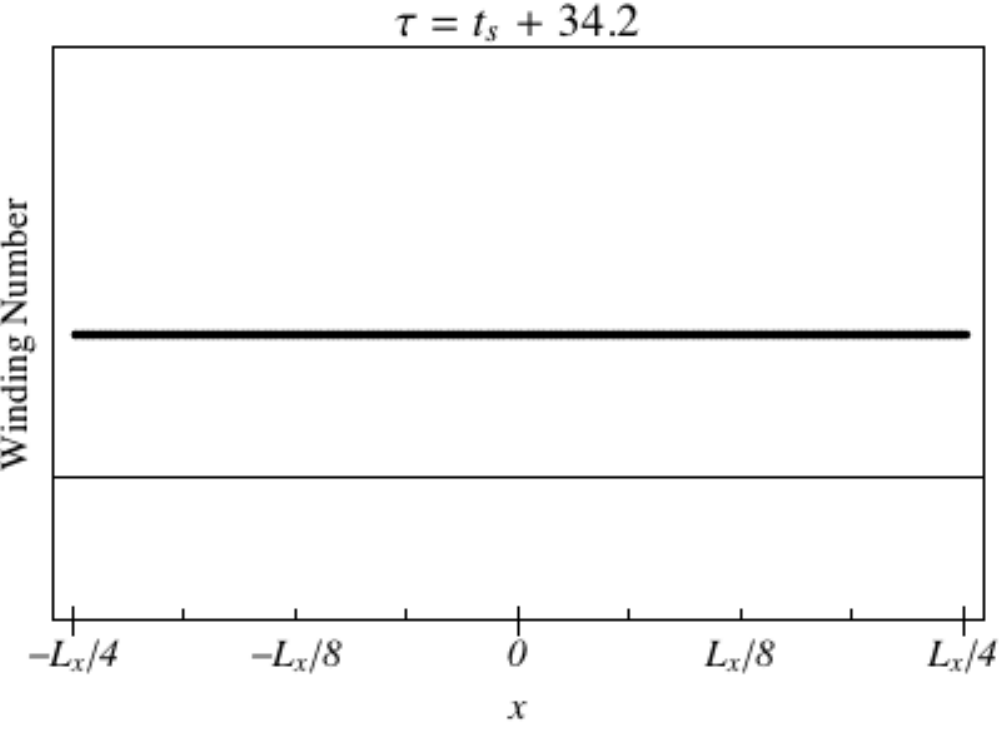} \\
  \includegraphics[width=2in]{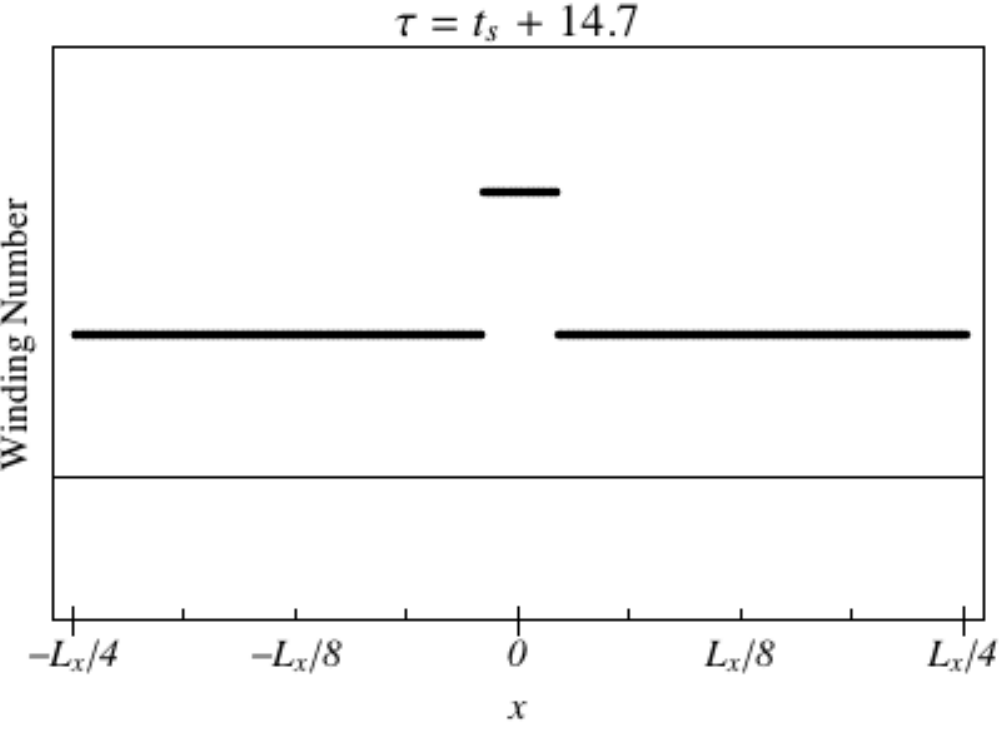}  
 \includegraphics[width=2in]{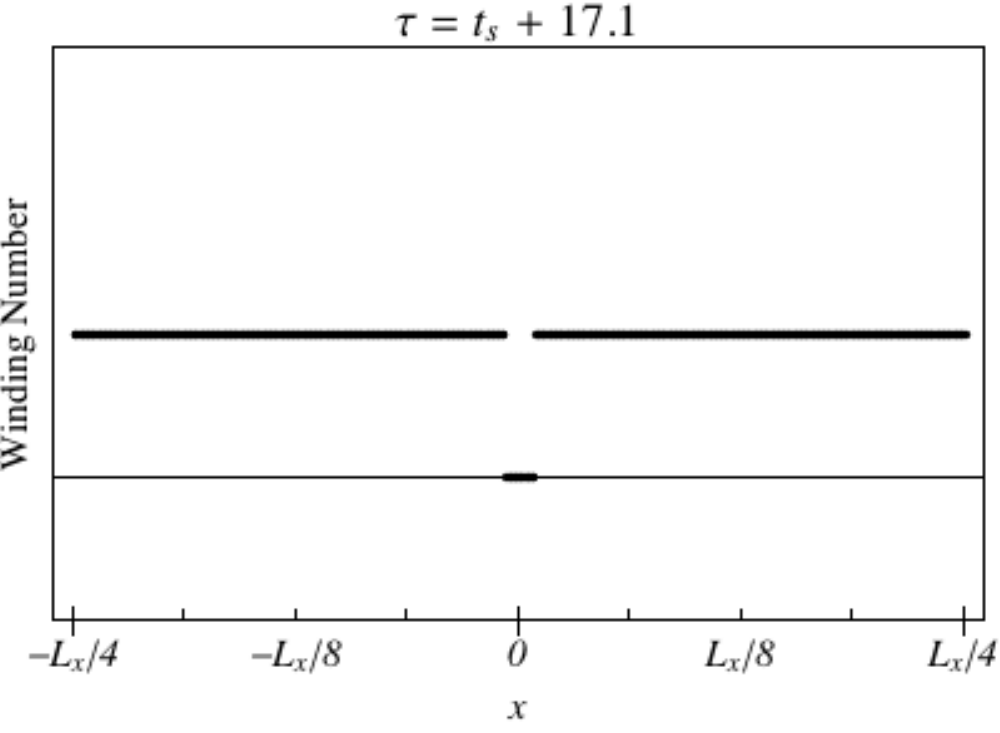}  \includegraphics[width=2in]{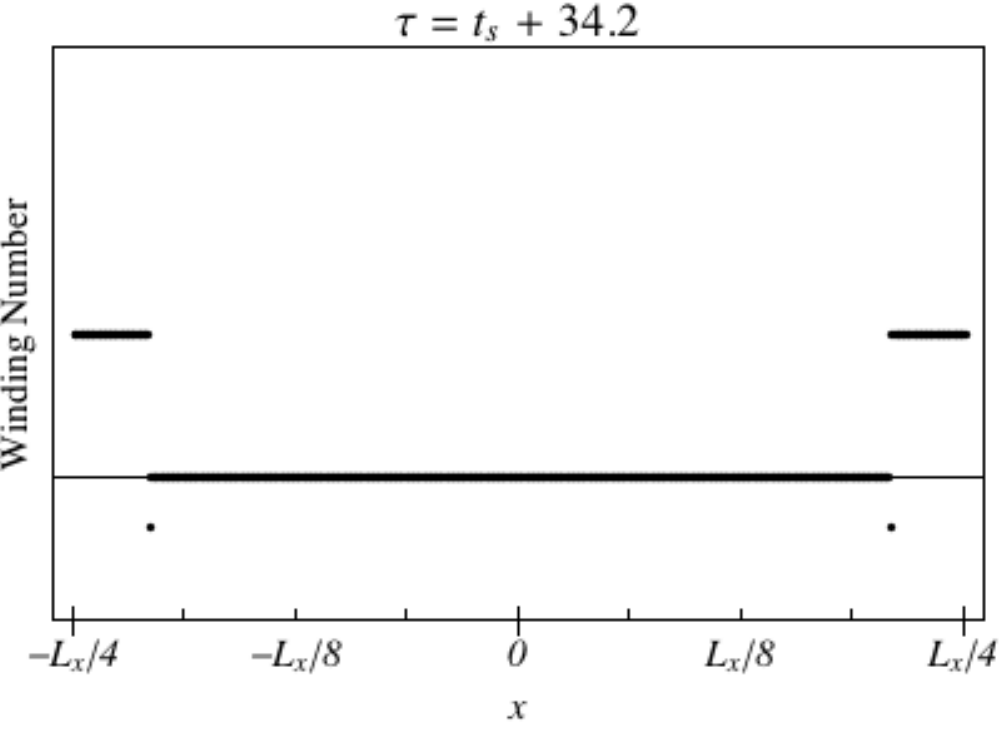} 
 \caption{The similar set of three snapshots of $w(x)$, for wall separation $\approx1.9 L$ and $\Delta\theta=0$ (top), $11\pi/16$ (middle) and $\pi$ (bottom).}
   \label{fig-1914}
\end{figure}

Next we keep the wall separation at $\approx1.9L$ and present $\Delta\theta=0,\ 11\pi/16$ and $\pi$.  As shown in Fig.~\ref{fig-1914}, the transition process at $0$ and $\pi$ are similar to each other, and much cleaner than Fig.~\ref{fig-5wind} where $\Delta\theta=9\pi/16$.  The non-transition result at $11\pi/16$ looks quite similar to Fig.~\ref{fig-4wind}.

\section{Discussion}

We have demonstrated numerical simulations of vortex collisions in the background with compact dimensions.  Since vortices are the natural boundaries for vacua with different winding number, we see the expected phenomena of classical transitions.  Our result agrees with the understanding from simpler models \cite{EasGib09,GibHui10} that more energetic collisions help to facilitate the transitions.  However, as explained in \cite{GibHui10} we should again advocate caution against a na\"ive energetic argument that larger collision energy helps to make two non-interacting vortices.  Since before the collision we have two fast approaching vortices, there is always enough energy to make two out-going ones.  Also as shown in Fig.~\ref{fig-5wind}, providing more initial energy to the collision may just create complicated final states, like more vortices or sourcing radiative modes, instead of directly helping the transition.  So one should think beyond the simple energetic argument. 

The more intriguing result is that against the intuition of \cite{BlaSch09}, larger vortex separation in the compact dimension actually reduces the chance of transit.  The near head-on collisions ($\Delta\theta$ close to $0$) almost always lead to classical transitions, and failures start to appear as $\Delta\theta$ increases.  So vortices ``missing'' each other actually does not always help.  More precisely, we cannot define ``missing'', since the vortices do not have finite sizes smaller than $L$, and there are always ``side-ways'' interactions.  The interaction basically transfers $x$ momentum to $y$ momentum. With less $x$ momentum, the vortices fail to fly apart.

This also explains why at $\Delta\theta=\pi$ we always get transitions, and they look very similar to the cases where $\Delta\theta=0$.  In both cases, the $Z_2$ symmetry forbids the vortices from acquiring any $y$ momentum.  So they tend to exit the collision without having actually collided.  Although $\pi$ happens to be the maximum separation of vortices in this model, we do not think the success of transitions should be attributed to this separation.

Finally, in the intermediate values of $\Delta\theta$ we observed interesting behavior.  For example the production of two extra vortices, as shown in Fig.\ref{fig-5wind}, is the most unexpected.  The short range interaction between vortices is an unexplored subject and we look forward to further developments: including more complicated compactified geometries.

\acknowledgments 
We thank Alan Guth, Alex Vilenkin, Jose Blanco-Pillado, and Ben Shlaer for a stimulating discussion and the Center for Theoretical Physics at the Massachusetts Institute of Technology for its hospitality during which time this discussion took place.  JTD and JTG are supported by the National Science Foundation, PHY-1068080, and a Cottrell College Science Award from the Research Corporation for Science Advancement.  The work of I-Sheng Yang is supported in part by the US Department of Energy, grant number DE-FG02-11ER41743.

\bibliographystyle{utcaps}
\bibliography{all}

\end{document}